\documentstyle[11pt,newpasp,twoside]{article}
\input psfig.sty
\index{instructions}
\index{guidelines}

\def\cxo{{\em Chandra}}
\def\hst{{\em HST}}
\def\vlt{{\em VLT}}
\def\xmm{{\em XMM}}

\def\edcomment#1{\iffalse\marginpar{\raggedright\sl#1\/}\else\relax\fi}
\reversemarginpar

\begin{document}
\title{HST and VLT observations of Pulsars and their Environments.}
 \author{R. P. Mignani}
\affil{ESO, Karl Schwarzschild Str.2, D85749 Garching}
\author{A. DeLuca, P.A. Caraveo}
\affil{IASF, v. Bassini 15, I-20133 Milan, Italy}

\begin{abstract}
The  state of  the art  of optical studies  of  Isolated Neutron Stars
(INSs) and their Pulsar Wind Nebulae (PWNe)  is reviewed. In addition,
results  obtained   from recent  \hst~   and  \vlt~  observations  are
presented and discussed. 
\end{abstract}

\section{Introduction}
The Crab  pulsar (PSR B0531+21)  was the  first isolated  neutron star
(INS) identified at  optical  wavelengths  (Cocke, Disney,  \&  Taylor
1968).  Almost a decade went by till another INS, the Vela pulsar (PSR
B0833-45), was detected  (Lasker 1976) and  confirmed  (Wallace et al.
1977).  The Vela pulsar, among the faintest objects known at that time
($V \sim 23.6$), was also the last INS detected by photograhic plates.
A  decade  later,  came  the first    CCD  detection of    the optical
counterpart of another INS, Geminga  (Bignami et al.  1987).   Spurred
by this  result and by  the  advent of  the new generation  telescopes
(e.g.  the ESO NTT), optical observations of INSs were carried on with
revived enthusiasm   and at the  beginning of  the 90s  yielded to the
detections of the counterparts of PSR B0540-69 (Caraveo et al.  1992),
readily confirmed by Shearer et al. (1994),  
PSR B0656+14 (Caraveo,   Bignami,  \& Mereghetti 1994a)   PSR B1509-58
(Caraveo, Mereghetti,  \& Bignami 1994b),  this  one later  revised by
Wagner \& Seifert (2000).  In the meantime, the identification
of Geminga was confirmed through the proper  motion measurement of its
counterpart  (Bignami, Caraveo,  \& Mereghetti  1993).   The launch of
\hst~ in 1990 increased the chances of detecting new INSs by providing
access to the near-UV window  through the short-wavelength sensitivity
of the Faint Object Camera (FOC).  Indeed, soon came the detections of
PSR B0950+08, PSR  B1929+10 (Pavlov, Stringfellow,  \& Cord\'ova 1996)
and PSR B1055-52 (Mignani,  Caraveo, \& Bignami 1997).  Likely optical
counterparts were also detected for  some of the  INSs singled out  in
the ROSAT data (a.k.a.  X-ray Dim INSs - XDINSs), namely RXJ 1856-3754
(Walter \& Matthews 1997) and RXJ 0720-3125  (Kulkarni \& van Kerkwijk
1998).  Thus, the number of INSs  observed in the optical/UV increased
by a factor four in just a decade.

\section{The identification status}
In recent years,  much effort was  concentrated on confirming  most of
the proposed counterparts.  Of course,  pursuing the identification by
searching for optical pulsations  at the radio/X-ray  period, although
ideally the best way, turned out to  be extremely difficult due to the
intrinsic  object   faintness.   Furthermore,   almost  none  of   the
middle-size/large   ground-based   telescopes   offered devices    for
high-resolution timing  and very few of  them were easily adaptable to
suited guest instruments.   Such problems prompted observes  to devise
alternative identification strategies.     The use of  proper   motion
measurements of the  candidate  counterpart, successfully experimented
in the case of Geminga (Bignami et al.  1993), proved to be a powerful
tool.  Indeed, proper motion  yielded the confirmation of the  optical
counterparts  of PSR B0656+14  (Mignani,   De Luca \& Caraveo   2000),
although a   marginal  detection  of  optical pulsations   did  exhist
(Shearer et  al.   1998),  of  RXJ  1856-3754 (Walter 2001),   of  PSR
B1929+10 (Mignani et  al.  2002) and  of RXJ 0720-3125 (Motch, Zavlin,
\& Haberl 2003), all but the last one achieved through high-resolution
\hst~  imaging.  In   the  meantime,   the  Space Telescope    Imaging
Spectrometer (STIS) on \hst, equipped with a UV-sensitive MAMA device,
took up the pathfinder task of the dismissed FOC.  In addition, with a
timing resolution of 125 $\mu$s, the  STIS made it possible to perform
timing observations of pulsars in   this spectral region.  Apart  from
the Crab pulsar  (Gull et al.,  these proceedings), four other pulsars
have  been observed to   date by the  STIS.  While  the  timing of PSR
B1929+10 has not provided  so far conclusive results (Mignani,  p.c.),
pulsations at the  known period  have  been clearly detected  from the
Vela pulsar and  Geminga (Romani \& Pavlov 2003)  as well  as from PSR
B0656+14 (Gull et al., these proceedings).  While  for the Vela pulsar
the STIS data  have provided the first measure  of its light  curve in
the  UV,  for PSR B0656+14  and  Geminga  detected pulsations provided
further confirmation of the identifications,  improving on the earlier
timing results of Shearer  et al.  (1997) and  Shearer et al.  (1998).
By exploiting the high   positional accuracy provided by  \cxo, likely
candidate optical counterparts have been also detected by the STIS for
two newly   discovered XDINSs,  RXJ 1308.6+2127   and  RXJ 1605.3+3249
(Kaplan, Kulkani,  \& van Kerkwijk 2002,2003).   The last entry is PSR
J0437-4715, the first  ms-pulsar  detected at optical/UV  wavelengths,
recently detected with the STIS  (Kargaltzev, Pavlov, \& Romani 2003).
Table  1    summarizes  the     current  (October  2003)    optical/UV
identification    status  of INSs.  As   it   is seen, owing  to their
intrinsic faintness, most INSs have been detected only thanks to their
proximity and small interstellar absorption.    Only the Crab  pulsar,
PSR  B1509-58 and PSR  B0540-69 have been detected  at more than 1 kpc
distance.    Looking at  Table    1 we see    that thanks  to the  new
telescopes/CCD  technologies, the INS   discovery rate  increased from
{\it one per decade} to almost {\it one per  year}.  In particular, we
note how the \hst~ contribution has been fundamental, providing nearly
all of the INS detections obtained in the  last 10 years i.e.  half of
the total.  As a matter of fact, so far \hst~ has clearly detected all
the INSs it was targetted  to.  Instead, the \vlt~ potentialities have
been only partially exploited in  the identification work, yielding so
far  only  to one  detection.   Finally,  proper motion  measurements,
although  requiring much  longer time   spans,  turned out  to be   as
efficient as timing in securing optical identifications of INSs.

\begin{table}[h]
\small
\begin{tabular}{l|lll|lll} \hline
{\em Name} & {\em Proposed} & {\em Confirmed} & {\em Evidence} & {\em mag} & {\em d(kpc)} & {\em $A_{V}$} \\  \hline
Crab           &1968&1969                          &TIM   &16.6&1.73&1.6\\
Vela           &1976&1977                          &TIM   &23.6&0.23&0.2\\
Geminga        &1987&1993/1998,{\em 2003}          &PM/TIM&25.5&0.16&0.07\\
B0540-69       &1992&1994                          &TIM   &22  &49.4&0.6\\
B0656+14       &1994&1997,{\em 2003}/{\em 2000}&TIM/PM&25  &0.29&0.09\\
B0950+08       &{\em 1996}&                                  &     &27.1&0.26&0.03\\
B1929+10       &{\em 1996}&{\em 2002}                        &PM    &25.6&0.33&0.15\\
B1055-52       &{\em 1997}&                                  &      &24.9&0.72&0.22\\
RXJ1856-3754   &{\em 1997}&{\em 2001}                        &PM    &25.7&0.14&0.12   \\    
J0720-3125     &1998&2003                                    &PM    &26.7&    &0.10   \\
B1509-58       &{\bf 2000}&                                  &      &25.7&4.18&5.2\\
RXJ1308.6+2127 &{\em 2002}&                                  &      &28.6&    &0.14\\
RXJ1605.3+3249 &{\em 2003}&                                  &      &26.8&    &0.06\\
J0437-4715     &{\em 2003}&                                  &  &    &0.14&0.11  \\ \hline
\end{tabular}

\caption{INSs identification status.   The first four columns give the
name,  the  publication year of  the counterpart  discovery and of its
unambigous   confirmation   (italics=    \hst,    bold=\vlt) and   the
identification   evidence,  i.e., either   pulsations  (TIM) or proper
motion (PM).  Columns five to  seven give the $V$-band magnitude  when
available (see Table 2), the distance in kpc, either obtained from the
DM     or      from          radio/optical      parallaxes        (see
$http://rsd-www.nrl.navy.mil/7213/lazio/ne\_model/$),    and       the
interstellar absorption $A_V$, as measured  directly or from the $N_H$
derived from the soft X-rays spectral fits.  } 
\end{table}

\section{New HST and VLT observations of INSs: The chase goes on}

Spurred by recent results, the search for  new optical counterparts of
INSs has been  pursued both with  the \hst~ and  the  \vlt.  After the
first-light  investigation of  PSR   B1706-44 (Mignani,   Caraveo,  \&
Bignami 1999),   the  \vlt~  observed  a  number  of  pulsars  and INS
candidates, although with not much luck so far.  \\ One of the obvious
targets was   the $\sim$ 1.7 Myrs  old,   nearby ($\sim  200$  pc) PSR
J0108-1431.  Although no potential     counterpart was found    at the
revised radio  position (Mignani, Manchester,   \& Pavlov  2003),  the
derived upper limits ($V \simeq 28$, $B\simeq  28.6$, $U \simeq 26.4$)
allowed to constrain the surface temperature of the neutron star to $T
< 8.8~ 10^4$~K (for $d$=200 pc and $R$ = 13 km), a  value in line with
the  expectations of  standard  cooling  models for   such an old  INS
(Mignani et al.  2003a).   The \vlt~ also observed  for the first time
ms-pulsars, all selected  according to their  $\dot E \approx 10^{33}$
ergs~s$^{-1}$,  X-ray emission, close  distance ($<$  500 pc) and  low
interstellar absorption.     No counterpart  was  identified  for  PSR
J2124$-$3358  ($U \ge  26$,  $B\ge 27.7$ and  $V  \ge27.8$; Mignani \&
Becker 2003) and PSR J0030+0451  ($B \ge 27.3$, $V \ge  27$ and $R \ge
27$; Koptsevich et al.   2003),  two ms-pulsars  very similar in  both
their  timing and X-ray emission.   For  PSR J0030+0451 (Koptsevich et
al.   2003),  and likely also for   PSR  J2124-3358 (Becker \& Mignani
2003), the optical flux upper limits  are well below the extrapolation
of the   non-thermal \xmm~ X-ray   spectrum.  In  case  of non-thermal
optical emission, this  would  imply a turnover in  the  optical/X-ray
spectrum.   The derived neutron   star  surface  temperatures  (13  km
radius) for PSR J2124-3358 (Mignani \& Becker 2003) and PSR J0030+0451
(Becker et al. 2003) are $\le 4.5  ~ 10^{5}$ K  and $\le 9 ~10^{5}$~K,
which are  above the value measured for  PSR J0437-4715 (Kargaltzev et
al.   2003).   Unconclusive    results were reported   from  shallower
observations of PSR J1024$-$0719 and  PSR J1744$-$1134 (Sutaria et al.
2003) for which  no spectral X-ray data  are available for comparison.
The \vlt~ observed also the 424 ms  pulsar 1E 1207$-$5209 in the young
($\sim$  7,000 years)  SNR G296.5+10.0  but  it  failed to detect  any
object  brighter than R$\sim$27.1 and V$\sim$27.3  within the $\sim$ 2
arcsec,  boresight-corrected,  \xmm~  error   circle  thus making this
apparently young object more similar  to middle-aged INSs (see also De
Luca  et al.  2003,  De  Luca et  al.  these  proceedings).  The 16 ms
X-ray pulsar PSR J0537-6910 has been  recently observed with the \hst,
taking advantage  of  the revised  and more  precise  ($\le 1$ arcsec)
\cxo~ position (Wang et al.   2001).  Thanks to the spatial resolution
and sensitivity of the recently  installed Advanced Camera for  Survey
(ACS)  it  was  possible to improve   on  the ground-based  results of
Mignani  et al.    (2000) and  resolve faint   stellar sources in  the
crowded core of the N157B SNR.  Multicolor imaging has pinpointed few,
new, potential counterparts characterized  by unusual colours (Mignani
et al.  2003b)  which  are being investigated through  timing analysis
with the \hst/STIS.  Finally, recent \hst~ and \vlt~ observations have
contributed  in clarifying the nature of  the puzzling Compact Central
Object  (CCO) 1E  161348-5051 in  the young ($\sim$   2,000 years) SNR
RCW103. Originally considered as   a  "good" INS candidate,  its  real
nature has  been debated after the  discovery of long term variations,
$\sim$ 6 hours periodicity and dips  of the X-ray  flux (see Becker \&
Aschenbach  and references  therein).   Deep IR observations with  the
\vlt~      have          detected       a  potential       counterpart
($J=22.3$,$H=19.6$,$K=18.5$) which has been  confirmed by  a follow-up
with the \hst.    Searches  for correlated  long  timescales  IR/X-ray
variability with \cxo~  (PI G.  Garmire)  as well  as  for a 6  hrs IR
periodicity have been unconclusive so far (Sanwal et al. 2003). If the
object is indeed  associated with the CCO, at  the  source distance it
could only be a low-mass star or a fossil disk.  It  could thus be the
first case of a LMXB identified at the center of a SNR.

\section{Optical emission properties of INSs}

The spectral database and the derived spectral properties for the INSs
in Table 1  are summarized in Table  2.  As  it  is seen, only in  six
cases (Crab, B0540-69,  Vela,  Geminga, J0437-4715 and RXJ  1856-3754)
the knowledge  of the  spectrum  can rely on  optical/UV spectroscopy.
Indeed, it is  basically   thanks to the    advent of the   10m  class
telescopes, like   the Keck and   the \vlt, that  spectroscopy  of the
faintest INSs became  possible.  For most INSs,  multicolor photometry
is still the only source of spectral information.   Only for four INSs
(Crab, Vela, B0656+14 and Geminga) the spectral coverage spans all the
way from  the IR to the  UV. For all  of them,  with the exception the
Crab, IR detections were provided for the  first time by the \hst~ and
the \vlt~.  This was  crucial to unveil the  presence of both  thermal
and non-thermal spectral  components, whose contributions are expected
to be markedly  different  in the  IR and  in  the UV.  For five  INSs
(B1509-58, B1055-52,  B1929+10, RXJ  1605.3+3249 and RXJ  1308.6+2127)
only  one,   two   passband    photometry  is   available     and  the
characterization of the optical spectrum is only tentative.

\begin{table}[t]
\small
\begin{tabular}{l|l|l|l|ll|l} \hline
{\em Name} & {\em Log $\tau$} & {\em Spec.}  & {\em Phot.} & {\em $\alpha$} & {\em T} &
{\em Comments} \\ \hline
Crab$^{1}$    &3.1 & 1100-9000 &UV,UBVRI,JHK& $-$0.11 & - & $PL_{o} < PL_{x}$    \\
1509-58$^{2}$ &3.19&           &R           &       &   & \\ 
0540-69$^{3}$ &3.22& 2500-5500 &UBVRI       & +0.2  &-  & $PL_{o} < PL_{x}$  \\ 
Vela$^{4}$    &4.05& 4500-8600 &UV,UBVRI,JH    & +0.12 &-  & $PL_{o} \sim PL_{x}$ \\ \hline 
0656+14$^{5}$ &5.05&           &UV,UBVRI,JHK& +0.45 &8.5& $PL_{o} \sim PL_{x}$ \\ 
              &    &           &            &       &   & $BB_{o} \sim BB_{x}$  \\
Geminga$^{5}$ &5.53& 3700-8000 &UV,UBVRI,JH & 0.8   &4.5& $PL_{o} < PL_{x}$  \\
              &    &           &            &       &   & $BB_{o} \sim BB_{x}$  \\
1055-52$^{6}$ &5.73&           &U           &       &   & \\ 
\hline
J0720-3125$^{7}$   & 6.4 &             & UV,UBVR  & $-$1.4 & 4$^{*}$ &$BB_{o} > BB_{x}$\\ 
1929+10$^{8}$ &6.49&           &UV,U        & +0.5  &-  & $PL_{o} < PL_{x}$ \\ 
0950+08$^{9}$ &7.24&           &U,BVI       & +0.65 &-  & $PL_{o} \sim PL_{x}$ \\  \hline 
0437-4715$^{10}$&9.2&1150-1700           &          & -     &1.0& $BB_{o} > BB_{x}$  \\ \hline 
J1856-3754$^{11}$   & ? & 3600-9000 & UV,UBV   &  -   &2.3&$BB_{o} > BB_{x}$ \\ 
J1605.3+3249$^{12}$ & ? &             & VR       &      &  \\ 
J1308.6+2127$^{13}$ & ? &             & V        &      &  \\ \hline
\end{tabular}
\vspace{0.1cm}

\noindent
$^{*}$ a distance of 300 pc is assumed

\noindent
$^{1}$Sollerman et al. (2000);
$^{2}$Wagner \& Seifert (2000);
$^{3}$Nasuti et al. (1997);
$^{4}$Shibanov et al. (2003);
$^{5}$Komarova et al. (2003);
$^{6}$Mignani et al. (1997);
$^{7}$Motch et al. (2003); 
$^{8}$Mignani et al. (2002);
$^{9}$Zharikov et al. (2003);
$^{10}$Kargaltzev et al. (2003);
$^{11}$van Kerkwijk \& Kulkarni (2001);
$^{12}$Kaplan et al. (2003);
$^{13}$Kaplan et al. (2002)

\caption{Spectral database for  the INSs in  Table 1, sorted  according to  their spin down  age $\tau$  (column  two) and grouped by age  decades. 
Column three and four give the
spectral  range (in  \AA) covered by  spectroscopy and  by photometry  (band-coded).  The spectral  index $\alpha$
and temperature  (in units  of $10^{5} K$)  of the  observed power-law
($F_{\nu} \sim \nu^{-\alpha}$ and blackbody components (hyphens
stand  for  non-detections)  are listed  in
columns five and six, respectively.   The last column tells wether the
 flux     of     the      PL/BB     optical
($PL_{o}$;$BB_{o}$)  components is higher  ($>$), lower  ($<$) or
comparable  ($\sim$)  to  the  optical extrapolation   of  the analogous X-rays components ($PL_{x}$;$BB_{x}$).
}
\end{table}

\noindent
Although
the  spectral  database  is  far  from being  complete,  some  general
patterns can  be recognized  in Table 2.   First of all,  the spectrum
grows in  complexity as  a function  of the INSs  age, passing  from a
single  power-law  (PL) to  a  composite  one  featuring both  PL  and
blackbody (BB) components.  The  underlying presence of a PL component
is  also  recognizable  from   the  correlation  between  the  optical
luminosity $L_{o}$ and  the rotational energy loss $\dot  E$ (see 
Kramer,  these  proceedings).
Although in some cases  the optical PL/BB
components  do match  the  extrapolation of  the analogous  components
observed in the X-ray domain, no general optical/X-ray correlation can
be  recognized. Thus, optical and X-ray emission mechanisms are not
always related to each other. The optical  emission  properties of  INSs can  be
summarized as follows for the different age class indentified in Table
2: \\
\noindent
(i) \underline{Young  INSs}: They are characterized by single
PL spectra, all with spectral index 
$\alpha_{\nu}>0$ apart from the Crab which shows evidence for a
spectral turnover in the optical/IR (Sollerman et
al. 2000). 
\\
(ii) \underline{Middle-aged INSs}: They feature composite  spectra  with  both  PL  and BB components, the former dominating
in the IR, the latter in the UV. In general they have spectral indexes  $\alpha_{\nu}
>  0$ which, at least for PSR B0656+14, seem to be steeper than in Young INSs. For both PSR B0656+14
and Geminga, the optical BB is consistent with the
extrapolation of the X-ray one and is likely produced from the whole
neutron star surface.
 For PSR B1055-52 nothing can be said but that the U-band flux is
consistent with the extrapolation of the X-ray PL. 
\\
(iii) \underline{Old INSs}: Both PSR B1929+10 and PSR B0950+08 feature single PL spectra with 
spectral indexes $\alpha_{\nu}
>  0$ and steeper than those of Middle-aged INSs. An
additional BB component might be present  but it can not constrained by the
present data.
\\
(iv) \underline{ms-pulsars}: The UV spectrum of PSR J0437-4715 is a BB, with a temperature higher than the one of the $> 20$ times younger PSR J0108-1431. For PSR J2124-3358 and PSR J0030+04651 (see \S3), the available flux upper limits  seem also to suggest a BB spectrum. 
\\
(v) \underline{XDINSs}: For RXJ 1857-3754 the spectrum is a BB and it is above the optical extrapolation of the X-ray one. On the other hand, the spectrum
of RXJ 0720-3125 seems to be composite, with a dominant BB component
(also above the optical extrapolation of the X-ray one) and a PL with
$\alpha_{\nu} <  0$.

\section{HST observations of Pulsar Wind Nebulae}

Although a few Pulsar Wind Nebulae (PWNe) have been detected in X-rays
by \cxo~ (Gaensler  et  al., these proceedings), optical  observations
are still limited to a few cases. Apart from the Crab (e.g., Hester et
al.  2002), \hst~ has  observed only  three  other young  pulsars with
PWNe. \\ The structure of the PSR B0540-69 PWN has  been resolved by the
\cxo/HRC (Gotthelf \& Wang   2000)  in three different  components:  a
point-like source  coincident with the  pulsar,  an elongated toroidal
structure around it and a  jet-like feature apparently protruding from
the pulsar.  In  the optical, a PWNe  is clearly detected in the \hst~
data of Caraveo et  al.  (2000) with a  very similar elongated pattern
and a shallow maximum southwest  of the pulsar.   Also the faint  jet,
marginally   visible  in   the \cxo~   image,   does have   an optical
counterpart. \\ The Vela PWN features   a brighter inner part  ($\approx
2'$), with two arcs, a jet, and a  counter-jet, symmetrical around the
pulsar's proper motion direction.  The inner PWN is then embedded into
an extended emission (outer PWN), surrounded  by a bean-shaped diffuse
nebula   ($\sim 2'\times  2'$) with  an   elongated region  of fainter
emission and a $100''$-long  outer jet southwest  and northwest of it,
respectively  (see  also Kalgartzev et   al.   these proceedings).  To
investigate  the  reality of  the     previously claimed optical   PWN
(\"Ogelman et al.  1989), Mignani et al.   (2003c) have compared \cxo~
and very deep \hst~ images of the field but no optical counterparts of
the   known X-ray features   could    be  identified  with   3$\sigma$
(extinction  corrected)    upper  limits   of    $\approx  27.9$   mag
arcsec$^{-2}$ and   $\approx   28.3$--27.8 mag  arcsec$^{-2}$  on  the
brightness of  the inner and  outer PWN, respectively.  By using wider
ESO {\sl NTT} and 2.2m images, Mignani et al.   (2003c) also set upper
limits   of {\bf $\approx  27.1$}   mag arcsec$^{-2}$  on the  optical
emission from the southwest extension  of the X-ray nebula.  While the
derived upper limits for   the inner/outer PWN   are not far from  the
extrapolation  of  the available X-ray/radio data,   the  ones for the
southwest extension  are at  least 3 orders  of  magnitudes above  the
expected value. For the PSR J0537-6910  PWN no conclusive results have
been obtained   so far. A  preliminar  search for  an optical  PWN was
performed by Wang et al.  (2001) using archived \hst~ observations but
they were  too shallow to set stringent constraints. Their results can now
be improved thanks to  recent,  much deeper, \hst/ACS observations  of
the field (Mignani et al. 2004).

\section{Conclusions}
                                                                               
INSs with either a secured or a likely optical counterpart amounts now
to 14, i.e., a  number comparable to those detected  in X-rays  in the
pre-ROSAT  era.  Thus, the  optical  branch, originally  confined to a
handful of representative cases,  is growing in importance occupying a
larger  and larger  niche   in the  multiwavelength  studies  of INSs.
Certainly, much work remains  to be done to reduce   the gap, both  in
terms of quality  and  quantity, with the   X-ray  domain.  More  deep
imaging observations are needed  to pinpoint new candidates, to  study
correlations between the  emission  properties (luminosities, spectra)
as a function  of the pulsars' parameters and   to search for  diffuse
emission structures (PWNe and  bow-shocks).  More  timing observations
are  required, both to secure faster  identifications and to provide a
broader characterization  of the  pulsars' lightcurves.  Finally, more
spectroscopy   is critical to better  define   the emission processes.
These are tasks both for the current 10 m class telescopes and for the
next generation  of extra-large telescopes   (Becker \& Mignani, these
proceedings). 

\acknowledgments

RPM is grateful to G.G. Pavlov for useful discussions and comments on
the manuscript and to S. Zane for anticipating her results before publication.

\end{document}